\documentstyle[preprint,epsfig]{aastex}
\shorttitle{The Homogeneity of Spheroids from Self-Consistent Hydrodynamical Simulations}
\shortauthors{Dom\'{\i}nguez-Tenreiro et al.}

\begin{document}

\title{The Lack of  Structural and Dynamical  Evolution of Elliptical Galaxies
since $z \sim 1.5$:
Clues from Self-Consistent Hydrodynamical
Simulations
}

\author{R. Dom\'{\i}nguez-Tenreiro
\footnote{Dpt.\ F\'{\i}sica Te\'orica C-XI,
Universidad Aut\'onoma de Madrid,
E-28049 Cantoblanco,
Madrid,
Spain;
rosa.dominguez@uam.es; jose.onnorbe@uam.es; hector.artal@uam.es; $^2$ Current address:
Dept.\ of Physics, Mahidol University, Bangkok 10400, Thailand; alex@astro.phys.sc.chula.ac.th; $^3$ Dpt.\ F\'{\i}sica y A.C., Universidad Miguel Hern\'andez,
E-03206 Elche, Alicante, Spain; arturo.serna@umh.es},
  J. O\~norbe$^{1}$,
A. S\'aiz$^{1, 2}$,
 H. Artal$^{1}$ and A. Serna$^{3}$
}

\begin{abstract}

We present the results of a study on the evolution of the
parameters that characterize the
structure and dynamics of the  relaxed elliptical-like objects
(ELOs) identified at redshifts $z=0, z=1$ and $z=1.5$ in a set of hydrodynamic,
self-consistent simulations operating in the context of a concordance
cosmological model.
The values of the stellar mass $M_{\rm bo}^{\rm star}$,
the stellar  half-mass radius $r_{\rm e, bo}^{\rm star}$, and
the mean square
  velocity for stars $\sigma_{\rm 3, bo}^{\rm star}$, have
  been measured in each ELO and found to populate, at any $z$, a flattened 
  ellipsoid close to a plane (the dynamical plane, DP). 
Our simulations  indicate that, at the intermediate $z$'s considered,
individual ELOs evolve, increasing their
$M_{\rm bo}^{\rm star}$,  $r_{\rm e, bo}^{\rm star}$, and
$\sigma_{\rm 3, bo}^{\rm star}$ parameters  as a consequence
of on-going mass assembly, but, nevertheless,
their DP is roughly preserved within its
scatter, in agreement with  observations of the fundamental plane (FP) at different $z$'s.
  We  briefly discuss how this lack
  of  significant dynamical and structural evolution
  in ELO {\it samples} arises, in terms 
  of the two different phases operating in the mass aggregation history
  of their dark matter halos.
  According to our simulations, most dissipation involved in ELO
  formation takes place at the early violent phase, causing the
  $M_{\rm bo}^{\rm star}$,  $r_{\rm e, bo}^{\rm star}$, and 
  $\sigma_{\rm 3, bo}^{\rm star}$ parameters to settle down to the DP
  and, moreover, the transformation of most of the available gas
  into stars.
  In the subsequent slow phase, ELO stellar mass growth preferentially
  occurs through non-dissipative processes, so that the DP is preserved
  and the ELO star formation rate  considerably decreases.
These results hint, for the first time, at a possible way of
explaining, in the context of cosmological  simulations,
different and apparently paradoxical observational results
for elliptical galaxies.
\end{abstract}

\keywords{ dark matter--- galaxies: elliptical and lenticular, cD--- galaxies: formation--- galaxies: evolution--- galaxies: fundamental parameters--- hydrodynamics
}

\section{Introduction}
\label{intro}

Among all galaxy families, elliptical galaxies (E's) are the simplest ones
and those that show the most precise regularities in the form of
relations among some of their observable parameters.
One of the most meaningful is the so-called fundamental plane
relation \citep[][]{Djo:87,Dres:87,Fab:87},
defined by their observed effective radius, $R_e^{\rm light}$,
mean surface brightness within that radius, $<I^{\rm light}>_e$,
and central line-of-sight velocity dispersion,
$\sigma_{\rm los, 0}$.
Analyses of the Sloan digital sky survey (SDSS)
sample of local elliptical galaxies
confirm previous results on the FP,
(see Bernardi et al. 2003a  and references therein). 
Studies  of the FP of early-type galaxies
up to $z \sim 1$  \citep{Vand:01,Vanv:03,Wuyts:04,Treu:05}
show that changes in the FP with $z$ can be described
in terms of the evolution of their  average stellar
mass-to-light ratio, $M^{\rm star}/L_B$, 
as predicted by  a  scenario of pure
luminosity evolution of their stellar populations, and
with no further need for any significant 
structural or dynamical evolution of the sample \cite[see, however,][]{Dise:05}.

Different authors interpret the tilt of the FP relative to the
virial relation as being caused by different assumptions concerning
the dependence of the dynamical mass-to-light ratios, $M_{\rm vir}/L$,
or the mass structure coefficients,
$c_{\rm M}^{\rm vir} = {G M_{\rm vir} \over 3 \sigma_{\rm los, 0}^{2} R_e^{\rm light}}$, on the mass scale, see discussion in \cite{Ono:05}.
A possibility is that $M_{\rm vir}/L$ grows systematically
with increasing  mass scale because the total dark-to-visible mass ratio,
$M_{\rm vir}/M^{\rm star}$, grows
\citep[as suggested, e.g., by][]{Cio:96,Pah:98,Capp:05}.
Otherwise, a dependence of $c_{\rm M}^{\rm vir}$ on the mass
scale could be caused by, among other possibilities, systematic differences in
the relative spatial distribution of the baryonic and dark mass
components of elliptical galaxies \citep{Cio:96}.
Recently, we have confirmed these possibilities \citet{Ono:05},
finding that the
samples of elliptical-like objects (ELOs) identified, at $z=0$,
in our fully-consistent cosmological hydrodynamical simulations
exhibit just such trends giving rise to {\it dynamical} planes.
They have also found that the physical origin of
these trends presumably lies in the
systematic decrease, with increasing ELO mass,
of the relative amount of
dissipation experienced by the baryonic mass component along ELO
mass assembly.
So, these results suggest that the
dissipative processes involved in the mass assembly of elliptical galaxies
 result in the tilt of the  FP
 relative to the virial plane.
 A consequence would be that the tilt of the FP 
  must be preserved during those time intervals when
 no significant amounts of dissipation
occur along E assembly.

 Concerning the epochs of dissipation along elliptical assembly,
 because star formation (SF) requires gas cooling, 
 the age distribution of the stellar populations of a given
elliptical galaxy should reflect the time structure of the dissipative
processes involved in its formation and assembly
(i.e., when the dissipation rate was high,  its time scale and so on).
These age distributions 
indicate that, in E galaxies,
most SF  occurred (1), at high $z$, (2) on short timescales,
and, moreover, (3) at higher $z$s and on shorter
timescales for increasing E mass 
\citep[they show age effects, 
see, e.g.,][]{Cal:03,Ber:03b,Thom:05}.
This would be the generic behavior of the dissipation rate history
for elliptical galaxies, should the assumption on its connection with their
SF rate history be correct.

Points 1 and 2 above are the basic ingredients of the so-called
{\it monolithic collapse scenario},
see details and discussion in \citet{Pee:02,Mat:03,Som:04}.
This scenario also explains another set of observational results on
E homogeneity, such as, for example,
(1), the lack of significant structural and
dynamical evolution of lens E galaxies, at least out to $z \sim 1$
\citep{Treu:04} (2), the lack of any strong structural evolution in the stellar mass-size
relation since $z \sim 3$ \citep{Tru:04,McI:05}
and (3), the confirmed existence of a population of  old,
relaxed, massive ($M^{\rm star} > 10^{11}$M$_{\odot}$) spheroidal galaxies at
intermediate redshifts \citep[$z \sim 1 - 2$,][]{Cim:02,Cim:04,Stan:04}
or even earlier \citep{Mob:05}.
 However, the monolithic scenario
  does not recover all the currently available observations on Es
    either. Important examples, suggesting that mergers at $z$s below
    $\sim 1.5 - 2$ could have played an important role in E assembly,
    are
      (1), the growth of the total stellar mass bound up in bright red galaxies
        by a factor of $\sim 2$ since $z=1$
\citep{Bell:04,Cons:05,Fon:04,Dro:04,Bund:05,Fab:05}, implying that
 the mass assembly of most Es continues below $z = 1$,
 (2), the signatures of merging observed by the moment out to intermediate $z$s
\citep{LeF:00,Pat:02,Cons:03,Cas:05},
in particular of major dissipationless mergers between spheroidal
	     galaxies \citep{Bell:05},
		     which translate into a relatively high  merger rate for
		       massive galaxies even below $z = 1$; and 
(3), the need for a young stellar component in some elliptical galaxies
\citep{Vand:03,Vanw:04}
 or, more particularly,
   the finding of blue cores
(i.e., recent  SF at the central regions)
and inverse colour gradients in 30\% - 40\% of
the spheroidal galaxies in some samples
out to $z \sim 1.2$ \citep[see][and references therein]{Men:04}.

The observational  results above  are paradoxical because they demand,
on the one hand, that  spheroids have  passively-evolving
stellar populations and an FP that preserves its tilt below $z \sim 1$
 or higher
 and, on the other hand, that  mass assembly is an on-going process
 for most of them and they still form some stars below these redshifts.
 In fact, one could think that, in principle, the tilt of the FP
 could be modified by these last processes, and so, 
 the preservation of  the FP tilt (or its evolution)
 is  a very important issue because
 it could encode a lot of relevant information on the dissipation rate history
 of elliptical samples, and, consequently, on the physical processes
 underlying E formation and evolution.
 In order to reconcile all these physical assumptions and 
 observational background within a 
 formation scenario, it is useful to study E assembly from simple
 physical principles and in the context of the global cosmological
 model. Fully-consistent gravo-hydrodynamic simulations are a
 very convenient tool for working out this problem \citep{Nav:94,SomL:03,Meza:03,Koba:05,Rome:05}.
 In fact, individual galaxy-like objects
 naturally appear as an output of the simulations, so that the parameters
 characterizing them can be measured and compared with observations.
 Concerning E assembly, the method has already proved to be useful.
 Apart from the physical origin of their FP \citep{Ono:05},
 the issue of age effects of their stellar populations
 has  been addressed in \citet{DSS:04},
   hereafter DSS04, where  clues on how these age effects arise
	 are given. Moreover, the structural and dynamical properties
	 of these ELOs have been found to be consistent with observations
	 \citep{Saiz:04,Ono:05}.
		 The next step  is to study 
		 whether the dynamical FP of ELO samples does not evolve
		 below $z \sim 1.5$ or so,
		 and whether this is or is  not consistent
		 with on-going mass assembly
		 as detected in real E samples, and if it is,
		 to try to understand how this arises with regard 
to the cooling rate, the SF rate  and the mass assembly
		 histories of  ELOs. 
		 These are the issues addressed in this Letter.

\section{The homogeneity of the elliptical population: clues from their assembly history}

We have run five hydrodynamical simulations
 in the context of a concordance cosmological
model \citep{Sper:03},
in which
 the normalization parameter
  has been taken slightly high, $\sigma_8 = 1.18$, as compared
   with the average fluctuations of 2dFGRS or SDSS galaxies
    \citep{Lah:02,Teg:03} to mimic an active
     region of the universe \citep{Evrard:90}.
Galaxy-like objects of different morphologies form in these
 simulations.
  ELOs have been identified as those objects having a prominent, 
        dynamically-relaxed stellar spheroidal component,
	            with no disks and a very low cold-gas
		    content. This stellar component
has typical sizes of no more than $\sim $ 10 - 40  kpc
(hereafter the {\it baryonic object} or "bo" scale) and it
is embedded in a halo of dark matter typically 10  times larger in size.
ELOs also have an extended corona of hot diffuse gas. 
We consider the whole sample of ELOs identified in each of the
five simulations at
$z=0, z=1$, and $z=1.5$ 
(samples E-Z0, E-Z1 and E-Z1.5, with 26, 24 and 16  ELOs, respectively)
and  analyze   the evolution of their mass and velocity distributions 
between $z=0$ and $z=1.5$ by comparing the ELOs in these three samples.
To run the simulations, we have used DEVA, a Lagrangian SPH-AP3M code
using particles to sample dark matter or baryonic (i.e., gaseous and stellar)
mass elements.
We refer the reader to \cite{Sern:03} for
details on the simulation technique
and to \cite{Saiz:04}
for details on its implementation in the runs we analyze here and
on  the  general sample
properties at $z=0$.
  All simulations started at a redshift $z_{\rm in} = 20$.
SF processes have been 
implemented, in the framework of the turbulent sequential
SF scenario \citep{Elme:02}, through a simple
parameterization
  that transforms   cold locally-collapsing gas,
  denser than a threshold density,
  $\rho_{\rm thres} $,
  into stars at a rate
  $d\rho_{\rm star}/dt = c_{\ast}  \rho_{\rm gas}/ t_g$,
  where  $t_g$ is a characteristic time-scale chosen
  to be equal to the maximum of the local gas-dynamical time
  and the local cooling time and   $c_{\ast}$ is the average
  SF efficiency at the scales resolved by the code.
  This is the empirical Kennicutt-Schmidt law \citep{Ken:98}.
  The five simulations share the same values for the SF parameters
  ($c_*$ = 0.3 and $\rho_{\rm thres} = 6.0 \times 10^{-25}$ gr cm$^{-3}$)
  and differ in the seed used to build up the initial conditions. 

To characterize the structural and dynamical properties of ELOs, we will
describe their three dimensional  distributions 
of mass and velocity
 through the three intrinsic (i.e., three-dimensional) parameters
(the stellar mass at the baryonic object scale, $M_{\rm bo}^{\rm star}$,
the stellar  half-mass radius
  at the same scale, $r_{\rm e, bo}^{\rm star}$,
  defined as that radius enclosing half the $M_{\rm bo}^{\rm star}$ mass,
  and the mean square
  velocity for stars, $\sigma_{\rm 3, bo}^{\rm star}$)
  whose observational projected counterparts
(the luminosity $L$, effective projected size  $R_{\rm e}^{\rm light}$, and stellar
central line of sight. velocity dispersion, $\sigma_{\rm los, 0}$) enter the
definition of the observed FP.
We use three dimensional variables rather than projected ones to
avoid projection effects.
To measure the structural and dynamical evolution of ELOs, we carry out
a principal component analysis of the E-Z0, E-Z1 and E-Z1.5 samples
in the  three dimensional variables
$E \equiv \log M_{\rm bo}^{\rm star}$,
$r \equiv  \log r_{\rm e, bo}^{\rm star}$ and
$v \equiv \log \sigma_{\rm 3, bo}^{\rm star}$
through their $3 \times 3$ correlation matrix $C$.
We have found that one of the eigenvalues of $C$ is, for the three ELO samples
analyzed, considerably smaller than the others, so that at any $z$ ELOs populate
a flattened ellipsoid close to a two-dimensional plane in 
$(E,r,v)$ space;
the observed FP is the observational manifestation of this 
dynamical plane (DP). 
The eigenvectors of $C$ indicate that the projection

\begin{equation}
E - \mbox{\~E}_{\rm z} = \alpha^{\rm 3D}_{\rm z} (r - \mbox{\~r}_{\rm z}) + \beta^{\rm 3D}_{\rm z} (v - \mbox{\~v}_{\rm z}),
\label{planeEq}
\end{equation}
where $\mbox{\~E}_{\rm z}, \mbox{\~r}_{\rm z}$ and $\mbox{\~v}_{\rm z}$ are
the mean values of the variables $E, r$ and $v$ at redshift $z$,
shows the DP viewed edge-on.
Table 1 gives the planes  eq. (\ref{planeEq})
for samples E-Z0, E-Z1 and E-Z1.5,
as well as their corresponding thicknesses, $\sigma_{\rm z, Erv}$, and
the distances
$d_{\rm z}$ of the point ($\mbox{\~E}_{\rm z}, \mbox{\~r}_{\rm z}, \mbox{\~v}_{\rm z}$)
to the E-Z0 plane.
We see that the sample averages
$\mbox{\~E}_{\rm z}, \mbox{\~r}_{\rm z}$ and $\mbox{\~v}_{\rm z}$
grow as $z$ decreases, but in any case 
$\mid d_{\rm z} \mid < \sigma_{\rm z, Erv}$,
so that they move {\it on} the E-Z0 plane within its rms scatter.
Moreover, only 8.4 \% and 0 \% of ELOs in E-Z1 and E-Z1.5 samples respectively,
are at distances greater than $\sigma_{\rm z=0, Erv}$ from the E-Z0 plane.
These results indicate that ELO evolution roughly preserves their DP.
The results in Table 1  
 strongly suggest that  the evolution shown by
 the FP of real elliptical galaxies
 is not the result of a dynamical or structural evolution of
 E samples,
  corroborating other 
  observational findings on elliptical homogeneity (see $\S$1). 
To try to understand these results,
we report on ELO assembly and
its effect on the DP preservation at intermediate and low redshift.

Our simulations indicate that 
ELOs are assembled out of mass elements that at high $z$ are
enclosed by those  overdense regions $R$
whose local coalescence length $L_c(t, R)$ \citep{Ver:94}
grows much faster than average,
and whose mass scale, $M_R$
(total mass enclosed by $R$)
is on the order of an E total (i.e., including
its halo) mass (see DSS04 and references therein).
The virial mass of the
ELO at low $z$, $M_{\rm vir}$,  is the sum of the masses of the
particles that belong to $R$  and are involved
into the ELO merger tree.
Analytical models, as well as N-body simulations indicate that two different 
phases operate  along  halo mass assembly: first, a violent fast one, in
which the mass aggregation rates are high,
and then, a  slower one, with lower mass aggregation rates
\citep{Wech:02,Zhao:03,SS:05}.
Our hydrodynamic simulations give us, for each ELO, its mass 
 aggregation track
(i.e., its mass  aggregation history along 
the main branch of the corresponding merger tree), and,
moreover, its dissipation rate \footnote{That is, the 
amount of cooling per time unit 
experienced by those gas particles that at $z=0$ form the 
ELO stellar component.} 
and star formation rate histories, among others;
these three histories are plotted
in Figure 1 for a typical ELO. 
We have found that the fast phase occurs through a 
multiclump collapse following turnaround of the overdense regions,
and it is characterized by fast head-on fusions experienced by
the nodes of the cellular structure these regions enclose,
resulting in strong shocks and  high cooling rates for their gaseous component,
and, at the same time, in strong and very fast star formation bursts (SFBs)
that transform most of the available cold  gas in $R$.
For the massive ELOs in this work, this happens between $z \sim 6$
and $z \sim 2.5$ and corresponds to a cold mode of gas aggregation, as in
\cite{Ker:05}.
Consequently, most of the dissipation involved in the mass assembly of
a given ELO occurs  in this violent early phase at high $z$;
moreover, its rate history is reflected by the SF rate history,
as illustrated in Figure 1.
This implies that the dynamical variables settle down to the
ELO  DP  at this early phase.
This plane is tilted relative to the virial plane as discussed in
$\S$1
see  \cite{Ono:05} and also  \cite{Rob:05} for pre-prepared mergers.

The slow phase comes after the multiclump collapse.
In this phase, the halo mass aggregation rate is low and the $M_{\rm vir}$
increment results from major mergers, minor mergers or continuous
accretion. Our cosmological simulations show that
the  fusion rates are generally low,
and that a strong SFB follows a major merger
only if enough gas is still available after the early violent phase.
This is very unlikely in any case, and it becomes more and more
unlikely as $M_{\rm vir}$ increases (see DSS04).
And so, these mergers imply only a modest amount of energy dissipation
or SF,
as the major merger in Figure 1 at $t/t_U \sim 0.66$  (where $t_U$ is the age if the universe) illustrates.
Note, however, that mergers play an important role in this slow phase:
$\sim $ 50\% of ELOs in the sample have experienced a major merger
event at $ 2 < z < 0$, which results in the increase of
the ELO mass content (see Fig. 1),
size $r$, and stellar mean square velocity $v$.
A  consequence of the lack of dissipation could be that the DP is
preserved  at the slow phase, and in fact,  we have found 
in our cosmological simulations that 
{\it dissipationless} merger events roughly 
preserve it see Table 1
\footnote{The preservation of the FP in pre-prepared dissipationless mergers
has already been studied by
\citet{Cap:95,GG:03,Nip:03} and \citet{Boy:05}.}.

Apart from ELO stellar mass growth following dissipationless mergers
in the slow phase,
our simulations indicate that
$M_{\rm bo}^{\rm star}$ can also increase as a consequence of newborn stars,
formed either (1), within the ELO itself from accreted gas or
gas coming in satellites,
that falls to the central regions before being turned into
stars (see the weak SFBs in Fig. 1 at $0.45 < t/t_U <0.68$)
or (2) through 
mergers, as the SB at $t/t_U \sim 0.68$ in Figure 1.
While the first implies quiescent modes 
of star formation \citep[see][]{Pap:05} and
could explain the blue cores observed in some relaxed
spheroids,  both could explain the  need for a young
 stellar population to fit some E spectra (see references above). 
Major merger events become less frequent as time elapses,
allowing for a higher fraction of relaxed spheroids.
Both, on-going  stellar mass assembly
(either through stellar mass aggregation or forming newborn
stars) and the decrease of the major
merger rate, imply an increase of 
the stellar mass density   contributed by
relaxed ELOs. In fact, we find that it
has changed by a factor of
2.1 between $z=1$ and $z=0$, consistent with 
empirical estimations (see $\S$1).
 
To sum up, the main result we report on is the homogeneity of the relaxed
ELO population with respect to $z$,
as measured through  the dynamical plane defined by their stellar masses,
three-dimensional sizes and mean square stellar velocities at different $z$s,
and, at the same time, the increase of the average
values of these parameters as time elapses.
The simulations also provide us with clues to how 
these evolutionary patterns arise
from  the physical 
processes involved in E formation,
namely, the plane's appearance at an early violent phase as a consequence
of E assembly out of gaseous material, with cooling  and on short timescales
(as in the monolithic collapse scenario),
and the plane's preservation during a later, slower phase,
where dissipationless merging plays an important role in
 stellar mass assembly (as in the hierarchical scenario
 of E formation).
This early gas consumption  of proto-ellipticals
also explains why  most of the stars of today elliptical galaxies
formed at high redshifts, while they are assembled later on
\citep[see][for similar conclusions from a semi-analytic model
of galaxy formation grafted to the {\it Millennium Simulation}]{deLu:05}.
Simulations also provide clues to why E homogeneity 
is consistent with the appearance of
blue cores as well as with  the increase of the stellar mass 
contributed by the E population as $z$ decreases.
We conclude that the
simulations provide a unified scenario whithin which most current
observations on elliptical galaxies can be interrelated.
This scenario shares some characteristics with previously proposed scenarios,
but it also has significant differences, mainly 
that stars form out of cold gas that had never been shock heated
at the halo virial temperature and then formed a disk,
as the conventional recipe for galaxy formation propounds
\citep[see discussion in][and references therein]{Bin:04,Ker:05}. 
To finish, let us note that the
scenario for elliptical galaxy  formation emerging from our simulations
has the advantage that their dark mass and gas aggregation histories
result from simple
physical laws acting on generic initial conditions (i.e.,
realizations of power spectra consistent
with CMB anisotropy data).

This work was partially supported by the MECD (Spain) through grants
 AYA-07468-C03-02 and AYA-07468-C03-03
  from the Plan Nacional de Astronomía y Astrofísica. We thank E. Salvador-Solé 
for discussions, and the Centro de Computaci\'on
  Cient\'{\i}fica (UAM) for computing facilities.
  A.S. thanks the Eurepean Union for Fonds Européens de Développement Regional
 (FEDER) financial support.

\bibliographystyle{mn}

\clearpage

\begin{figure}
\includegraphics[width=\textwidth]{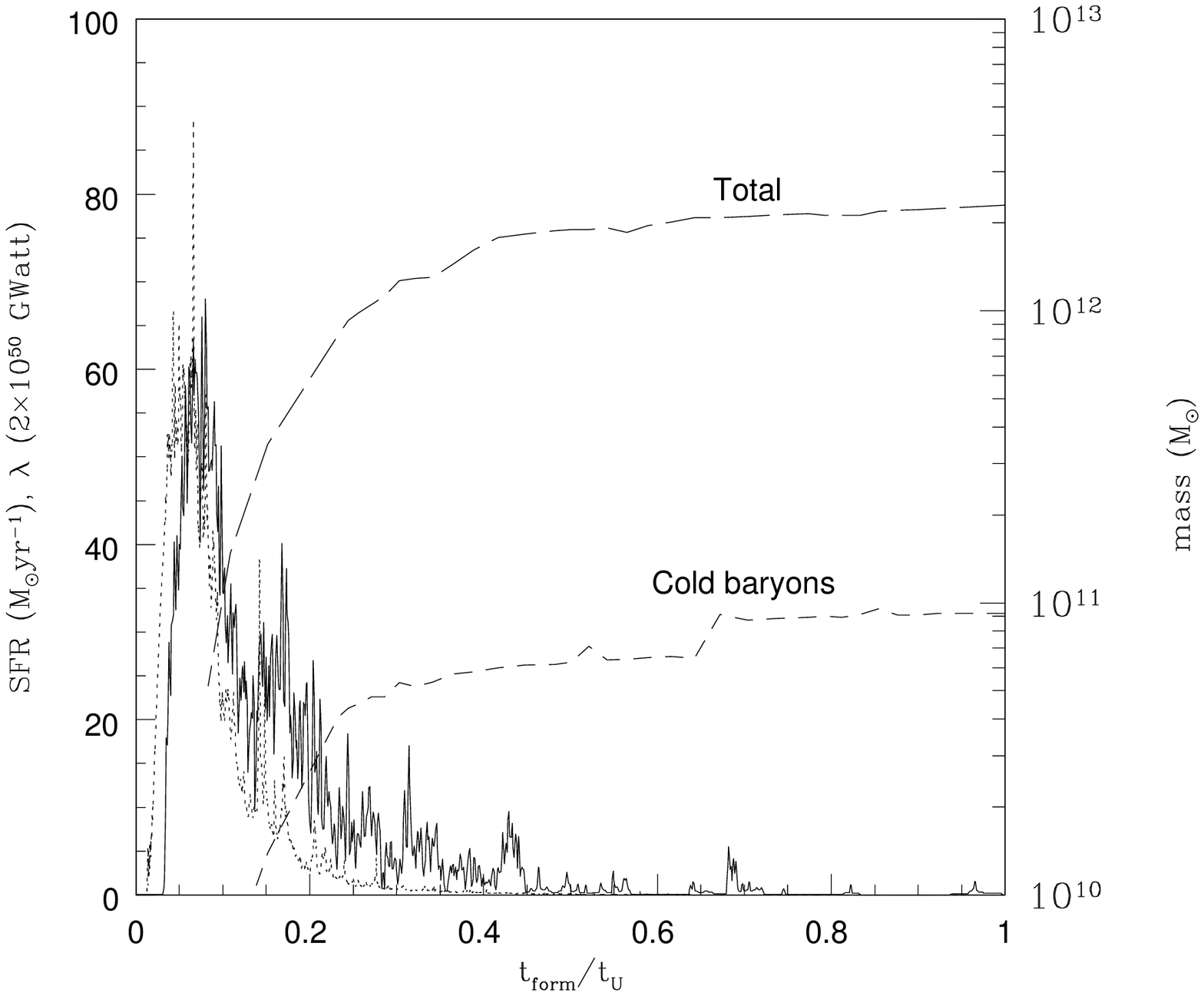}
\caption{ Cooling rate history $\lambda$ (dotted line) and the  star
formation rate history (solid line) of a typical ELO in the simulations.
 We also plot its mass aggregation tracks, both for  cold baryons
 (i.e., stars and cold gas) and total mass. 
The fast (left) and slow (right) phases of mass aggregation are clearly shown.
}
\label{Kplots}
\end{figure}
\newpage

\begin{table}
\caption{}
\begin{center}
\footnotesize
\begin{tabular}{llllllll}
\hline
Sample &  $\mbox{\~E}_{\rm z}$& $\mbox{\~r}_{\rm z}$& $\mbox{\~v}_{\rm z}$& $\alpha_{\rm z}^{\rm 3D}$& $\beta_{\rm z}^{\rm 3D}$& $\sigma_{\rm z, Erv}$ & d$_{\rm z}$  \\
\hline
\hline
E-Z0&  11.0& 0.75& 2.34& 0.43& 2.07& .011&  \\
E-Z1&  10.8& 0.52& 2.30& 0.25& 2.10& .011& .008  \\
E-Z1.5&  10.9& 0.49& 2.33& 0.31& 2.01& .013& .009  \\
\hline
\hline

\end{tabular}
\end{center}
\end{table}

\end{document}